\documentclass[review]{elsarticle}
\PassOptionsToPackage{hyphens}{url}\usepackage{hyperref}
\usepackage{lineno}
\usepackage{subfigure}
\usepackage{listings,xcolor}
\usepackage{placeins}

\colorlet{punct}{red!60!black}
\definecolor{delim}{RGB}{20,105,176}
\colorlet{numb}{magenta!60!black}
\definecolor{background}{HTML}{FAFAFA}

\lstdefinelanguage{json}{
    basicstyle=\normalfont\ttfamily,
    stepnumber=1,
    numbersep=8pt,
    showstringspaces=false,
    breaklines=true,
    backgroundcolor=\color{background},
    frame=lines,
    literate=
     *{0}{{{\color{numb}0}}}{1}
      {1}{{{\color{numb}1}}}{1}
      {2}{{{\color{numb}2}}}{1}
      {3}{{{\color{numb}3}}}{1}
      {4}{{{\color{numb}4}}}{1}
      {5}{{{\color{numb}5}}}{1}
      {6}{{{\color{numb}6}}}{1}
      {7}{{{\color{numb}7}}}{1}
      {8}{{{\color{numb}8}}}{1}
      {9}{{{\color{numb}9}}}{1}
      {:}{{{\color{punct}{:}}}}{1}
      {,}{{{\color{punct}{,}}}}{1}
      {\{}{{{\color{delim}{\{}}}}{1}
      {\}}{{{\color{delim}{\}}}}}{1}
      {[}{{{\color{delim}{[}}}}{1}
      {]}{{{\color{delim}{]}}}}{1},
}

\usepackage{amssymb}

\newcounter{bla}

\usepackage{amsmath}
\usepackage{fancyvrb}

\journal{Computer Physics Communications}

\begin{document}

\begin{frontmatter}



\title{DECal, a Python tool for the efficiency calculation  of thermal neutron detectors based on thin-film converters}


\author[a]{\'Alvaro Carmona Bas\'a\~nez\corref{author}}
\author[a]{Kalliopi Kanaki}
\author[a]{Francesco Piscitelli}

\cortext[author] {Corresponding author.\\\textit{E-mail address:} alvaro.carmonabasanez@esss.se}
\address[a]{European Spallation Source ERIC, P.O. Box 176, SE-221 00 Lund, Sweden} 
%\address[b]{Second Address}

\begin{abstract}
The Detector Efficiency Calculator (DECal) is a series of Python functions and tools designed
to analytically calculate, visualise and optimise the detection
efficiency of thermal neutron detectors, which are based on thin-film
converters. The implementation presented in this article concerns
$^{10}$B-based detectors in particular.  The code can be run via a graphical user interface, as well as via the command line. The source
code is openly available
to interested users via a GitHub repository.
\end{abstract}

%\begin{keyword}
%keyword1; keyword2; keyword3; etc.

%\end{keyword}

\end{frontmatter}

% \linenumbers

% Computer program descriptions should contain the following
% PROGRAM SUMMARY.

{\bf PROGRAM SUMMARY}
  %Delete as appropriate.

\begin{small}
\noindent
{\em Program Title:} DECal                                         \\
{\em Licensing provisions: Free for non-commercial use (as per terms in LICENSE file)}                                   \\
{\em Programming language:}                                  Python Version 2.7 and 3.3 \\
{\em Operating system:}  OS X (preferred) and Linux    \\
{\em Version:} 1.0.0
%{\em Supplementary material:}                                 \\

{\em Keywords:}  \\
\texttt{Detector efficiency calculator}, neutron scattering, Boron-10,
neutron detector, efficiency, thin film, converter, Python, PyQt5, \\

{\em Nature of problem:}\\

Implementation of a Python-based tool to calculate and optimise detection
efficiency for thin-film thermal neutron detectors via a graphical user interface and a command line interface.
  %Provide any additional comments here.

%% \begin{thebibliography}{0}
%% \bibitem{1}Reference 1         % This list should only contain those items referenced in the                 
%% \bibitem{2}Reference 2         % Program Summary section.   
%% \bibitem{3}Reference 3         % Type references in text as [1], [2], etc.
%%                                % This list is different from the bibliography at the end of 
%%                                % the Long Write-Up.
%% \end{thebibliography}
%% * Items marked with an asterisk are only required for new versions
%% of programs previously published in the CPC Program Library.\\
\end{small}

\let\clearpage\relax
\section{Introduction}
%% This manuscript describes a neutron detector configurator and
%% efficiency calculator tool which consists of a series of python
%% scripts and a graphical user interface (GUI) designed to calculate,
%% visualize, optimize and compare the efficiency of B10 based neutron
%% detectors or, more generally, detector based on neutron converter
%% layers. The source code is available in github and is open source so
%% the users can use and modify it at their will.

The European Spallation Source (ESS) ERIC~\cite{esstdr} is a joint
European organisation committed to the construction and operation of
the world's leading facility for research using thermal neutrons. The ESS is
designed to be the world's brightest neutron source and the
instantaneous neutron flux on detectors at ESS full power will be without
precedent. In general, neutron scattering facilities are going toward
higher fluxes and this translates into higher demands on instrument
and detector performance: higher counting rate capability, better
timing and finer spatial resolution are requested among others.

The current thermal neutron detector technology is reaching fundamental limits and
most of the neutron sources in the world, including ESS, are pushing
the development of their detector technologies~\cite{HE3S_cooper,HE3S_gebauer,HE3S_karl,HE3S_kirstein},
to tackle the increased flux available and the scarcity of $\mathrm{^3He}$, the so-called ``Helium-3
crisis''~\cite{HE3S_kouzes,HE3S_kramer,HE3S_shea}. $\mathrm{^{10}B}$
along with the $\mathrm{^3He}$ and $\mathrm{^6Li}$ isotopes are the
main actors in thermal neutron detection due to their large neutron
absorption cross sections~\cite{illblue}. Due to its favourable properties,
$\mathrm{^3He}$ (a rare isotope of $\mathrm{He}$) has been dominating
thermal neutron detection for decades.

Nowadays, the importance of solid conversion layers is increasing as this technology appears to be a viable and promising alternative to
$\mathrm{^3He}$. Recently, high-quality, low-cost production of square
metres of $\mathrm{^{10}B_4C}$-coated substrates~\cite{B4C_carina,Schmidt2016,HOGLUND201514} became possible. The
detection efficiency of a single thin layer of $\mathrm{^{10}B_4C}$ is limited
to a few percent at thermal neutron energies compared to the very high
efficiencies of $\mathrm{^3He}$-based detectors. The
Multi-Blade~\cite{MIO_HERE,MIO_MB2014,MIO_MB2017}, the Jalousie
detector~\cite{DET_jalousie,DET_Jalousie3}, the
A1CLD~\cite{DET_kampmannA1CLDp,DET_kampmannA1CLDp2}
and many others~\cite{MPGD_CrociRate,MPGD_GEMcroci}, are examples of
the detector developments which exploit solid neutron converters
operated at a grazing angle in order to increase the detection
efficiency. A different way to increase the efficiency is to arrange
many layers in sequence. Examples of detector developments arranging
several layers are the
Multi-Grid~~\cite{MG_2017,MG_andersen,MG_IN6tests,MG_joni,MG_patent},
CASCADE~\cite{MPGD_KleinCASCADE, kleinphd} and many
others~\cite{DET_stefanescu1,DET_stefanescu2,MPGD_pfeiffer2015,STRAW_lacy2,STRAW_lacy2002,STRAW_lacy2006,STRAW_lacy2011,STRAW_lacy2012,STRAW_lacy2013}.

The difference between the physical processes of a gas-based (such as
$\mathrm{^3He}$) and a solid-converter-based detector (such as
$\mathrm{^{10}B_4C}$) is described in detail in~\cite{mcgregor,piscitelli2013};
a detector based on solid converter has several parameters that can be
tuned to increase the detection efficiency. A simple model of single
and multi-layer thermal neutron detectors is explored mainly
analytically to help optimize the design in different
circumstances. Several theorems are deduced that help guide the
design. Using powerful simulation software has the advantage of
including many effects and potentially results with high accuracy~\cite{icnskelly}. On
the other hand it does not always give the insight an equation can
deliver. The equations in~\cite{piscitelli2013, piscitelli2013bis} and
the tool described in this manuscript focus on
$\mathrm{^{10}B_4C}$-based thermal neutron detectors but can be extended to any
detector based on thin-film converters.

%% MAYBE OMIT THE EQUATION
%% In particular, when a neutron interacts with a $\mathrm{^{10}}B$ nucleus, the reaction releases the following fragments: 
%% \begin{equation}
%% \begin{aligned}
%% \mathrm{n} + \, \mathrm{^{10}B} \rightarrow \mathrm{^7Li^*} + \alpha  &\rightarrow \mathrm{^7Li} + \alpha +  \gamma \, (478\,\mathrm{KeV}) &+  2.31 \,MeV  &\,\,(94\%)\\  
%%                                                              &\rightarrow  \mathrm{^7Li} + \alpha                           &+ 2.79 \,MeV   &\,\,(6\%)
%% \end{aligned}
%% \end{equation}

The theoretical details of the calculations which this software
implements in Python are described in~\cite{piscitelli2013,piscitelli2013bis} and more specifically
in the Ph.D.\,thesis~\cite{FPThesis}. The authors explain a series of
equations to calculate the efficiency of detectors based on thin film
neutron converters given the specific geometry of the detector and the
neutron beam characteristics.

%\textcolor{red}{+++ rephrase last paragraph to pass the ball to the
%  next section, maybe move it to next part}

%This tool allows to analytically calculate the efficiency of a detector. A detector can be made of a single and many layers; the latter can be hit by the incoming neutrons at any angle. Note that the angle affects the efficiency, as well as the neutron wavelength. A detector can hold many layers of same thickness or the thickness of the neutron converter can be adapted for different layers. The thicknesses can be optimized accordingly for a single neutron wavelength or for a distribution of neutron wavelengths which is a scenario closer to reality in a neuron scattering instrument. 

%HERE WE DO NOT EXMAPLIN THE EQUATIONS BUT ALVARO IN THE TEXT WILL REFER TO THEM GIVING A 1 SENTENCE EXPLANATION 

%PROBABLY NOT THIS PART!
%\subsection { B10 detector designs}
%\subsection {Solid converter leads to geometry whose efficiency can be calculated analytically}

\section {Detector geometry and neutron beam configuration}

A detector can be made of a single or multiple layers. A layer consists
of a substrate material, usually Aluminium, on which the $^{10}$B$_4$C
converter is coated, either on one or both sides. A double-coated layer is referred to as blade. The number of converter layers, their thickness and their composition
matter. In addition, a detector can
contain layers of the same or varying converter thickness. This
parameter can be optimised accordingly for a single neutron
wavelength or for a distribution of neutron wavelengths, a scenario
which is closer to reality in a neuron scattering instrument. The material and thickness of the substrate are not considered
in the calculations presented here and will be the topic of a future improvement.

\begin{figure}[!h]
  \centering
  \includegraphics[width=1\columnwidth]{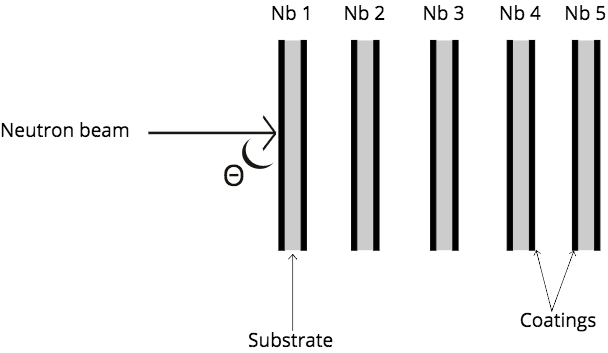}
  \caption{Depiction of detector geometry and neutron beam
    arrangement. The beam hits the parallel blades with an incident angle $\Theta$. The blades consist of a substrate on which the converter
    is coated on both sides.}
  \label{geometry_beam}
\end{figure}

The incoming neutrons can enter this geometrical
arrangement at an angle or perpendicularly, depending on the needs of
the application (see Fig.~\ref{geometry_beam}). Both the neutron
incident angle and its wavelength (energy) affect the
efficiency and thus enter the respective calculation function.% These are the two
                                % beam parameters that enter the efficiency calculation function.

%% When it comes to the detector, the number of converter layers, their thickness and their composition
%% matter. The material and thickness of the substrate are not considered
%% in the calculations presented here and will be the topic of a future improvement. In addition, a detector can
%% contain layers of the same or varying converter thickness. This
%% parameter can be optimized accordingly for a single neutron
%% wavelength or for a distribution of neutron wavelengths, a scenario
%% which is closer to reality in a neuron scattering instrument.

Fig.~\ref{geometry_beam} summarizes all parameters that impact the
detection efficiency. The neutron beam hits the detector with an
incident angle $\Theta$. A numbered series of layers follow. A layer is called back-scattering when neutrons are incident from the
gas-converter interface and the escaping particles are emitted backwards into the gas volume; it is called a transmission layer when neutrons are incident from the substrate-converter interface and
the escaping fragments are emitted in the forward direction in the
sensitive volume. In Fig.~\ref{geometry_beam} the detection efficiency
of every blade is the sum of the back-scattering and transmission layer
efficiencies, as the substrate holds two converter layers, one in back-scattering mode and one in
transmission mode.
%The beam and geometry parameters listed above enter the calculation
%of detection efficiency. 

%% \begin{figure}[!h]
%%   \centering
%%     \subfigure[Blade explatnation sketch]{\includegraphics[width=0.75\columnwidth]{bladeSketch}
%%     \label{fig:g4_al}
%%   } \caption{\label{fig:al} . Variables definition for a back-scattering and transmission layer calculations.}
%% \end{figure}

%\subsection {Description of the neutron beam}

%% \subsection {Explanation of equations for calculation of efficiency }
%% As seen in the thesis \cite{piscitelli2013} 
%% there are different equations to calculate efficiency depending on the parameters of the detector and the parameters of the neutron beam. As the equations get more complex they can be used also for more simpler calculations, for example the multi layer detector with a distribution of neutron wavelengths can be used for calculating a single layer detector with a monochromatic wavelength.

%\subsection {Example given of a geometry, efficiency calculation and optimisation}

\FloatBarrier

\section{Software overview}
DECal has been developed as a
cross-platform software for \textit{Mac OS X}  (version 10.11 and
higher) and \texttt{Linux}  (tested on CentOS 7 and Ubuntu 17.4). It has been written
in Python 3. The objective of this project has been to
develop an open source tool that allows the user to calculate the
detection efficiency of a $^{10}$B-based multi-blade neutron
detector and optimise the respective geometry parameters.

The software features allow the user to set the parameters of the neutron beam and detector with intuitive visualisation features. % The functions used to calculate the theoretical efficiency of a B10 based tetector were available in Matlab  %TODO link to matlab code by FP
%, the functions have been rewritten in Python. 
The package containing the efficiency calculation functions is distributed via Python Package Index (pip).   % TODO add reference A computer with dual core cpu is sufficient to make the calculations. %review
 %%TODO fix this and place where it is relevant

The configurations can be exported in \textit{JSON} format, and this \textit{JSON} file can be imported with graphical interface as well as in the functions of the library as input. 

%REview change the order: say hoy we keep data and then say that thats why there is not DB

%TODO where to place this sentence:  The graphical user interface has been developed using \textit{Pyqt} as interface design tool. It allows to add Matplotlib~\cite{matplotlib} plots easily and make use of it's features more easily.

\subsection{Software availability}

The software is divided in two different repositories. DECal is available in \href{https://github.com/DetectorEfficiencyCalculator/dg_efficiencyCalculator}{https://github.com/DetectorEfficiencyCalculator/dg\_efficiencyCalculator}
\href{https://github.com/alvcarmona/neutronDetectorEffFunctions}{In the  repository
  https://github.com/alvcarmona/neutronDetectorEffFunctions} the users can find the library that holds the calculation functions. It is also available via easy install via the following command:
 \begin{verbatim}
 $   pip install neutron_detector_eff_functions
\end{verbatim}
  %The GUI repository is this one \href{https://github.com/DetectorEfficiencyCalculator/dg_efficiencyCalculator} {(https://github.com/DetectorEfficiencyCalculator/dg_efficiencyCalculator)} %information about the project can be found in the project's official webpage \href{https://detectorefficiencycalculator.herokuapp.com/}  {website} 
  %TODO Show links with highlight style

\subsection{Requirements and dependencies}
This is a list of hardware and software requirements for the users to
access all the features of the tool:
\begin{itemize}
\item \textit{OS X }$\ge$10.10, \textit{Linux Ubuntu 17.4 and CentOS 7} are the OS where the software has been tested
\item Multi-core CPU
\item 1200$\times$610 screen resolution or higher
\item Python v2.7 or 3.3 and higher
\item Qt5 for the DECal application
\end{itemize}
The Python libraries available via pip needed to execute the DECal software are included in the requirements.txt file found in the repository. These are some of them:
\begin{itemize}
  \item SciPy \cite{scipy}
  \item NumPy \cite{numpy}
  \item Matplotlib $\ge$ 1.5.0 \cite{matplotlib}
  \item PySide $\ge$  1.2.4 
  \item PyQt5 (not available via pip in some systems)

\end{itemize}
For more information about the environment needed the user can refer
to the
\href{https://github.com/DetectorEfficiencyCalculator/dg_efficiencyCalculator/blob/master/requirements.txt}{requirements.txt file}.

\FloatBarrier
\section {Software Design}
 %TODO -  Write this better? Explain how it works briefly justify the c

%% \begin{figure}[!h]
%%   \centering
%%     \subfigure[Blade explanation sketch]{\includegraphics[width=1\columnwidth]{configurationSketch.pdf}
%%     \label{fig:g4_al}
%%   } \caption{\label{fig:al} . Variables definition for a back-scattering and transmission layer calculations.}
%% \end{figure}

The classes used to represent the Detector and its parts are the following: \textbf{Detector}, \textbf{Blade}, \textbf{B10}. The \textbf{Detector} class is the object that represents a detector configuration and has the plot functions.  %has the following attributes, converter, name, wavelength, angle, threshold, single, metadata and blades. %The following functions: calculate_eff(), calculate ranges(), calculate sigma(), plot_blade_figure(), plot_blade_figure_single(), plot_thick_vs_eff(), plot_wave_vs_eff()
%TODO Introduce the design
To deal with the construction of the  \textbf{Detector} entity with the different parameters there is a class that is built with the Builder pattern~\cite{builderp}. The builder pattern builds a complex object using simple objects and using a step by step approach. The builder function builds a detector entity depending on the parameters given.\\
%Efftools holds the efficiency calculation functions and the metadata functions for plotting. This functions have been directly translated from the original functions written in Matlab %TODO reference to repository or something

\subsection {File structure}
%TODO dont say it's simple

Script functions are in \textit{scripts.py}. In this folder there are other basic Python files like the requirements file and the license.
 The \textit{exports} folder contains some examples of exported neutron configurations and wavelengths. 
 % The main package is called \textbf{neutron_detector_eff_functions}, here hare the next source files:
 The \textbf{Detector} class is written in \textit{Detector.py}. %has the following attributes, converter, name, wavelength, angle, threshold, single, metadata and blades.
\textbf{Efftools} holds the core efficiency calculation functions and the metadata functions for plotting. The entire tool is built around efficiency4boron and efficiency2particles functions. These functions are the ones used to calculate the theoretical efficiency of a  neutron detector using B10 as converter.\\
\begin{verbatim}

============
neutronDetectorEffFunctions library file structure
============

README.md
LICENSE.txt
neutron_detector_eff_functions/
    Aluminium.py
    B10.py
    Blade.py
    Converter.py
    Detector.py
    Detector_meta.py
data/
    Aluminium/
        AlCrossSect_(n,g).py
    B10/
        10B4C220/
            IONIZ_Alpha06.py
            IONIZ_Alpha94.py
            IONIZ_Li06.py
            IONIZ_Li94.py
        10B4C224/
            IONIZ_Linkoping_Alpha06.py
            IONIZ_Linkoping_Alpha94.py
            IONIZ_Linkoping_Li06.py
            IONIZ_Linkoping_Li94.py
        B10CrossSect_(n,a).py

efftools.py
scripts.py # examples of use
requirements.txt
setup.cfg
setup.py
tests/
    Detector_test.py
    test_B10.py


\end{verbatim}

\iffalse
============
Detector Efficiency Calculator file structure
============

efficiencyCalculator/     
    exports/
        waves/
            2gaussdistr.txt
        Detector 118effVsDepth.txt 
        Detector 11effVsDepth.txt
        PolicromPolidconfig.json
        detector1.json
    tests/
        B10_test.py
        Detector_test.py
    detectorDialgog.py
    detectorDialogTab.ui
    detectorform.ui
    efficiencyCalculator.py
    efficiencyCalculator.spec
    efficiencyMAinwindow.ui
    efficiencyMAinwindow2.ui
.gitignore  
.hgignore
LICENSE
README.txt
launch.py   # Script to launch GUI application
launch.spec
requirements.txt
setup.py
\fi

%TODO explain data and the rest of mentioned things

The \textit{data} subpackage contains the files the application uses to calculate the cross-section of different converter materials. In the current version of the software there are two available configurations of B10. These files have been generated with a tool called SRIM  (The Stopping and Range of Ions in Matter)~\cite{SRIM}. This data is the stopping power of the materials. The user can find the description of the materials in the header lines of the files efficiencyCalculator/data/B10/. %TODO check this link
This is an example of this information: \\
\begin{verbatim}
============= TARGET MATERIAL ======================================
Layer 1 : Layer 1
Layer Width = 50000.E+00 A ;
  Layer # 1- Density = 12.74E22 atoms/cm3 = 2.207 g/cm3
  Layer # 1-  B = 77.6 Atomic Percent = 74.4 Mass Percent
  Layer # 1-  C = 20   Atomic Percent = 23.0 Mass Percent
  Layer # 1-  B = 2.4  Atomic Percent = 2.53 Mass Percent
====================================================================
 Total Ions calculated =000999.00
===============================================================
  Ionization Energy Units are  >>>>  eV /(Angstrom-Ion)  <<<<  
===============================================================
\end{verbatim}

\begin{figure}[!h]
  \centering
  \includegraphics[width=1.25\columnwidth]{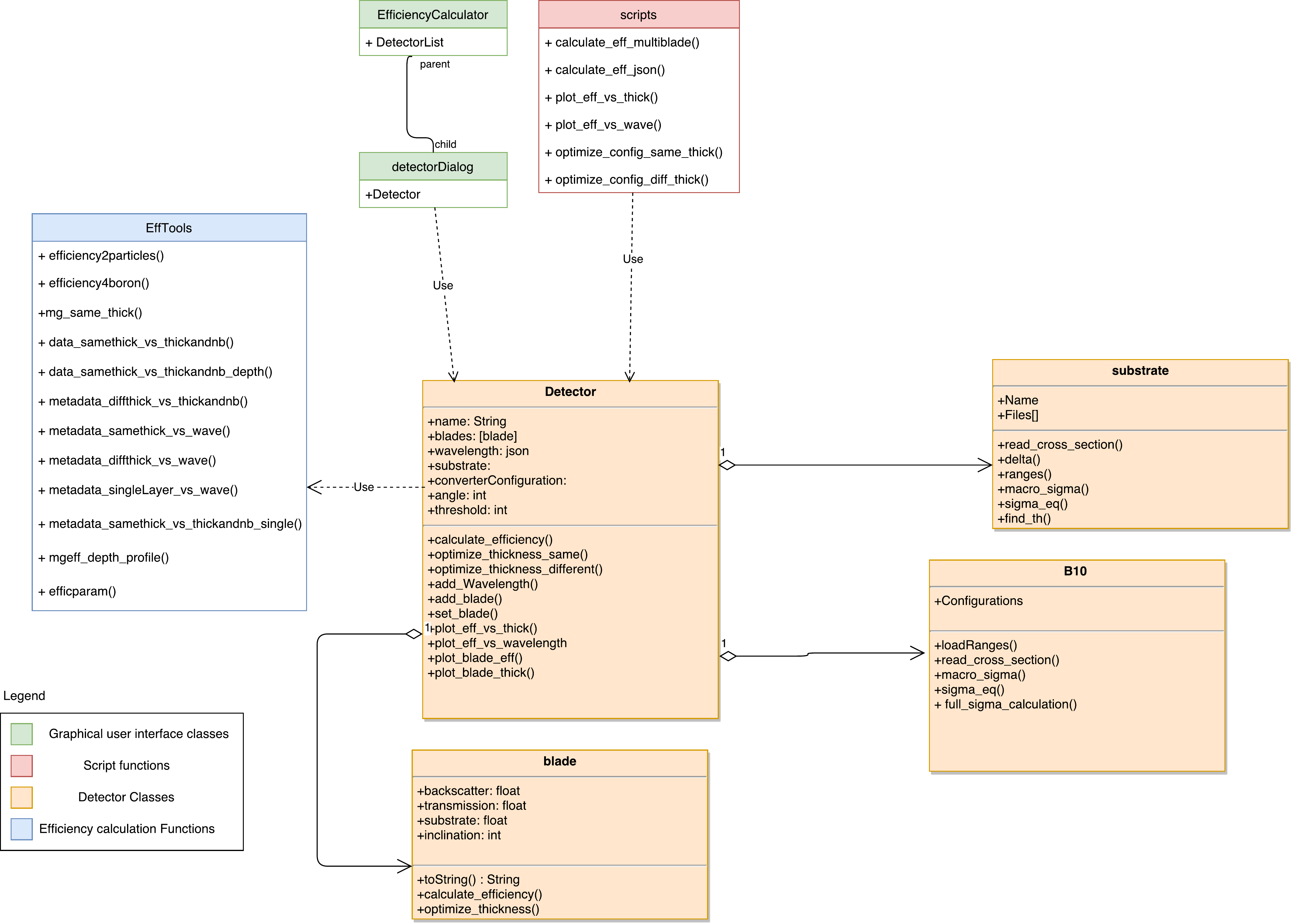}
  \caption{Class diagram of the DECal project.}
  \label{classdiagram}
\end{figure}

%\subsection {Classes}
%TODO fix number reference to figure
Fig.~\ref{classdiagram} represents the class diagram of the DECal project.

%TODO place this figure correctly
%The detector
%\\
%Blade
%\\
%B10
%\\
%DetectorDialog
%\\
%efficiencyCalculator
%\\

\FloatBarrier
\section {High level functions}
A set of functions has been developed for simplifying the use of the
classes included in the project. The functions can be called using the \textit{
scripts.py} file located in the root folder of the library. The
explanations listed in the following serve documentation purposes.
\begin{verbatim}
calculate_eff_multiblade(nb, converterThickness, substrateThickness,
 wavelength, angle, threshold, single, converter)
\end{verbatim}
 The function returns a list of two positions. The first contains a
   list of the efficiency for each blade in depth order (the last is
   the deepest) and the second value is the total efficiency of the
   detector. This is an example of usage: 

\begin{verbatim}
calculate_eff_multiblade(10, 1, 0, [[1.8,100]], 90, 100, False, 
'10B4C 2.24g/cm3')
\end{verbatim}
The function call returns the efficiency of a neutron detector
configuration of 10 blades with a converter thickness of 1 micrometre,
0 micrometres of substrate, a neutron wavelength of 1.8~\AA\,
(monoenergetic), 90 degrees of incident neutron angle, an applied threshold of 100 keV on the total energy deposition from the ionized
conversion products, double-coated layers and a $^{10}$B$_4$C
converter with a density of 2.24~g/cm$^3$ \cite{B4C_carina}.

\begin{verbatim}
calculate_eff_json(path)
\end{verbatim}
The function calculates the efficiency of a configuration loaded from a \textit{JSON} file. The user can generate this configuration via the graphical user interface.

\begin{figure}[!ht]
  \centering
    \includegraphics[width=0.9\columnwidth]{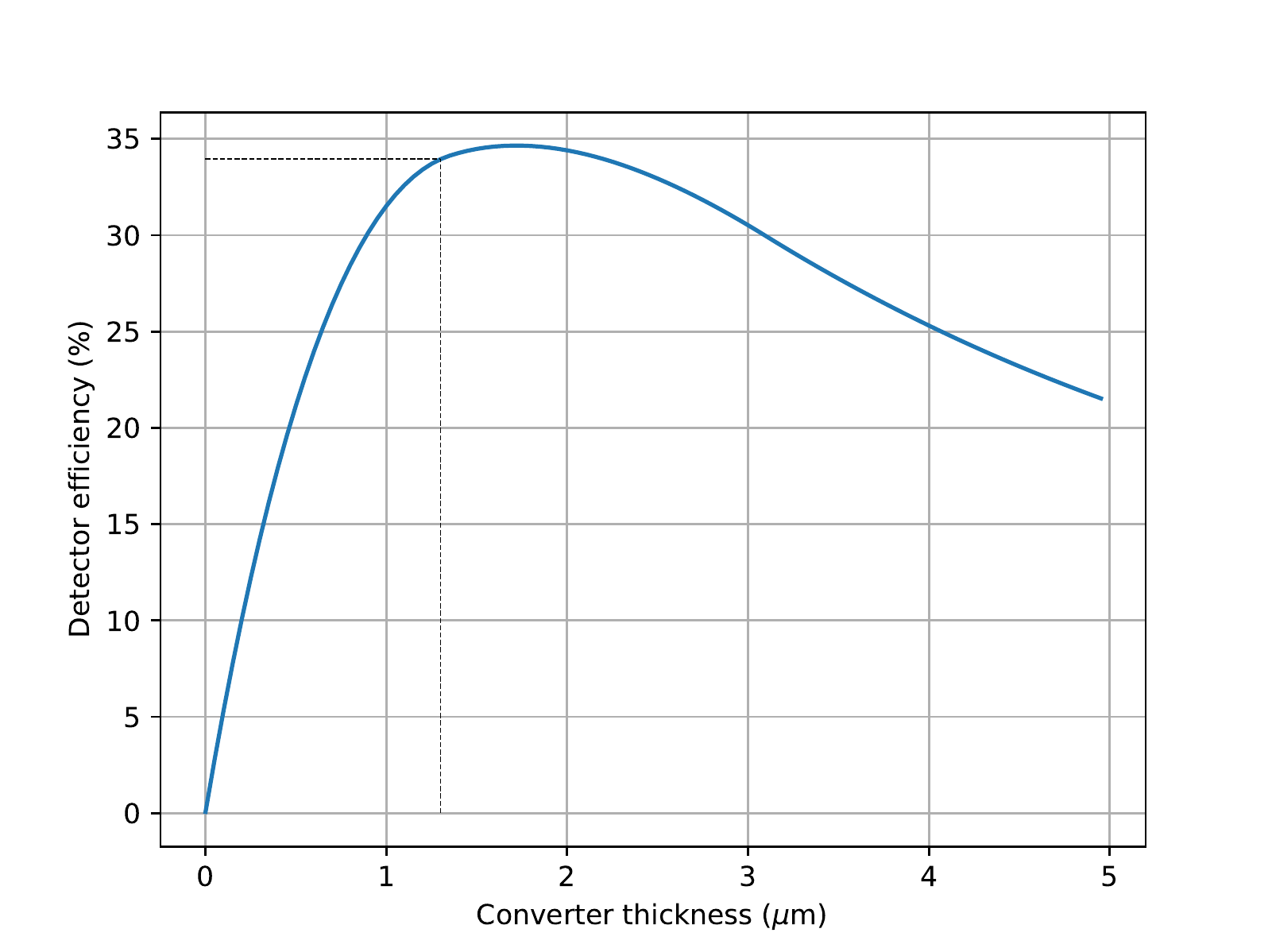}
    \caption{Detector efficiency vs.\,converter thickness for a set of
      default beam and detector parameters, e.g.\,1.8\AA\,~wavelength, 1.3~$\mu$m
      converter thickness, read from a typical \textit{JSON} file in GitHub.} 
    \label{asd}
\end{figure}

\begin{verbatim}
plot_eff_vs_thick(path)
\end{verbatim}
This function plots the detector efficiency as a function of different
converter thicknesses. The argument is the path to a \textit{JSON}
configuration file. A typical result is presented in Fig.~\ref{asd}.
%%TODO add description of the configuration parameters to be able to reproduce it

\begin{figure}[!h]
  \centering
  \includegraphics[width=0.9\columnwidth]{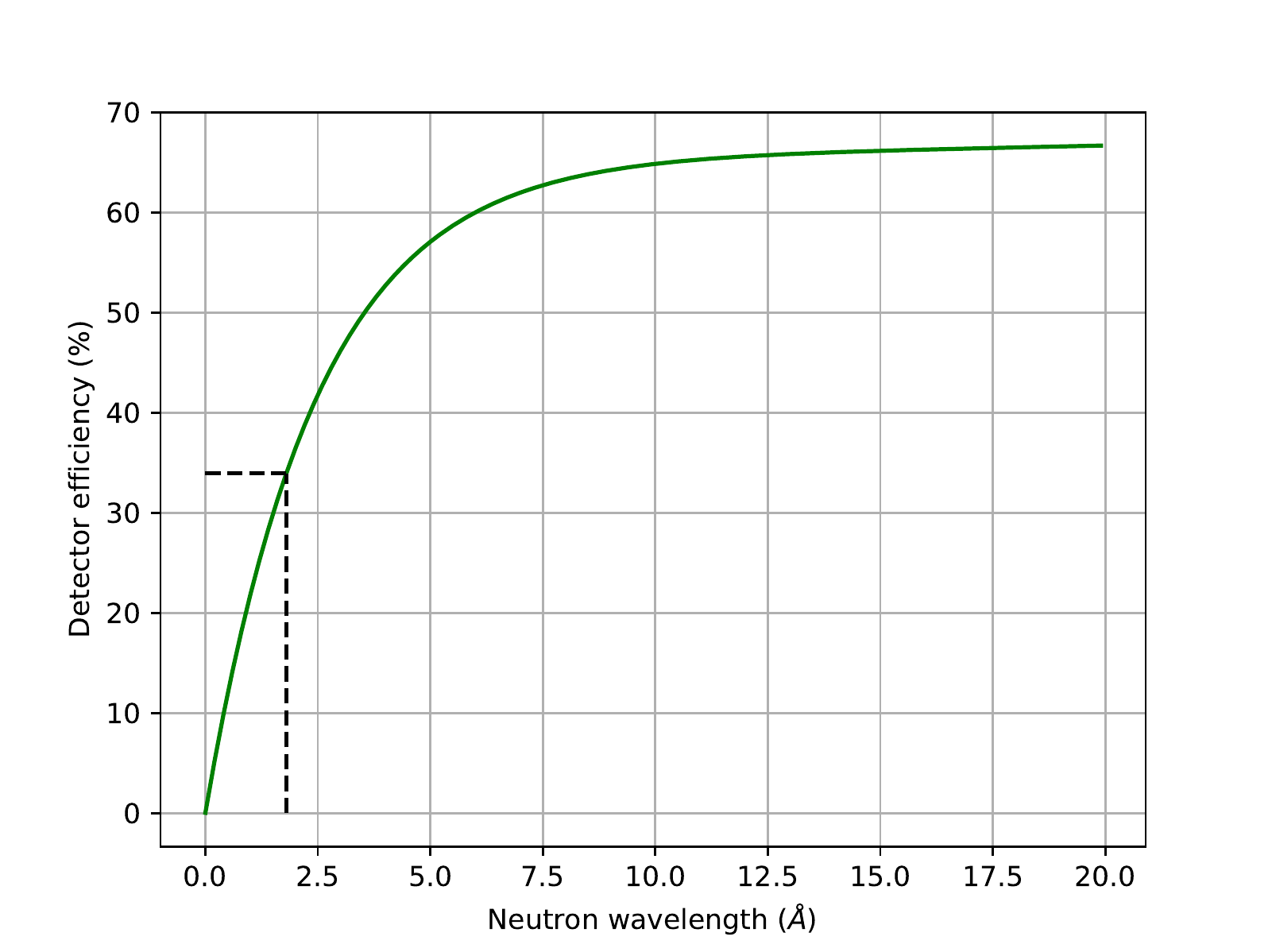}
  \caption{Detector efficiency vs.\,neutron wavelength, produced from a typical \textit{JSON} file in GitHub.} 
  \label{effvswavelength}
\end{figure}

\begin{verbatim}
plot_eff_vs_wave(path)
\end{verbatim}
Providing a path to a \textit{JSON} file as parameter, this function
plots the efficiency as a function of a monochromatic neutron
wavelength. The resulting plot is displayed in Fig.~\ref{effvswavelength}.

\begin{verbatim}
optimize_config_same_thick(originPath, destinyPath) 
\end{verbatim}
% Reference to json structure
The function returns a list of converter thicknesses optimised to
maximise detector efficiency, with the condition that all converter
thicknesses are identical. \textit{originPath} points to a
configuration from a \textit{JSON} file. \textit{destinyPath} is the
destination where the new configuration is saved in \textit{JSON} format.

for each blade the same converter
thickness is applied on both substrate sides.
\begin{verbatim}
optimize_config_diff_thick(originPath, destinyPath)
\end{verbatim}
Same as before but with the condition that different blades can have
different converter thicknesses.

\FloatBarrier

\section {Graphical user interface tool}
\begin{figure}[!tb]
  \centering
  \includegraphics[width=0.7\columnwidth]{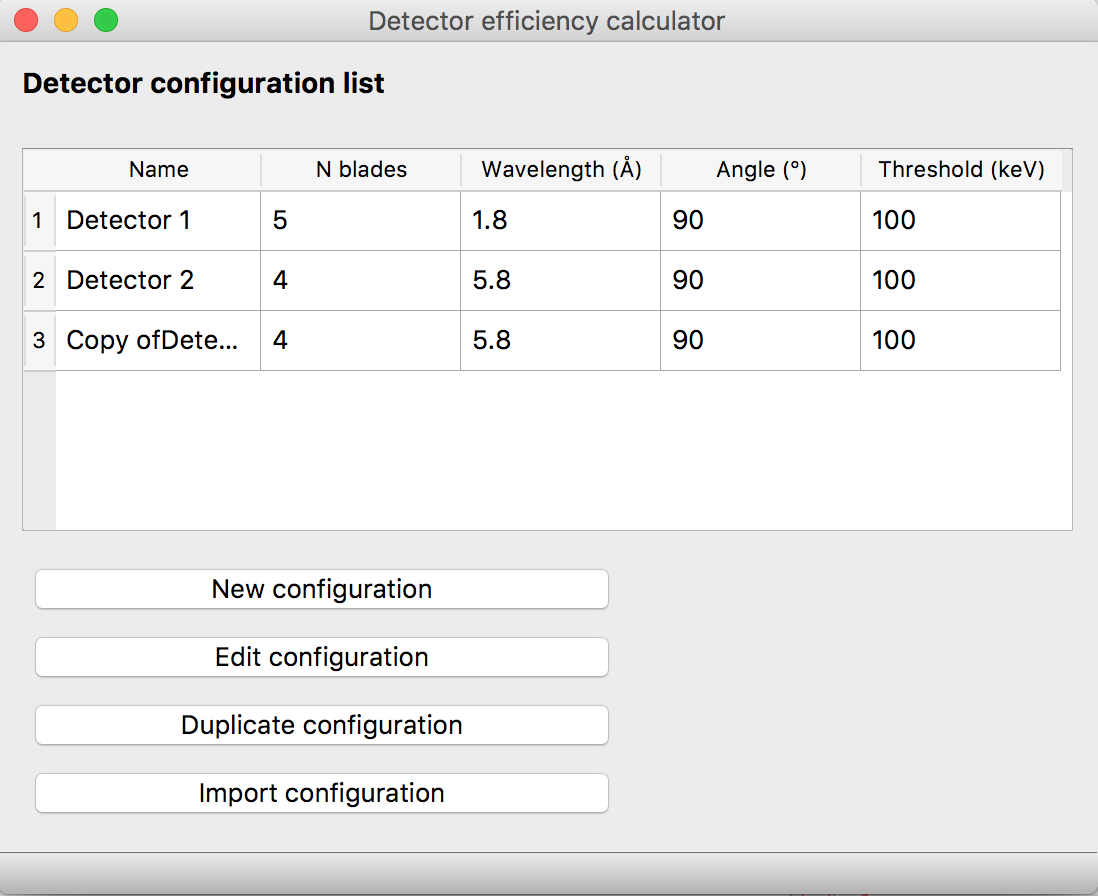}
  \caption{Main GUI window, where detector configurations are listed. New configurations can be created from here or existing ones can be edited, duplicated and imported.} 
  \label{guimainwindow}
\end{figure}

\begin{figure}[!b]
  \centering
  \includegraphics[width=0.9\columnwidth]{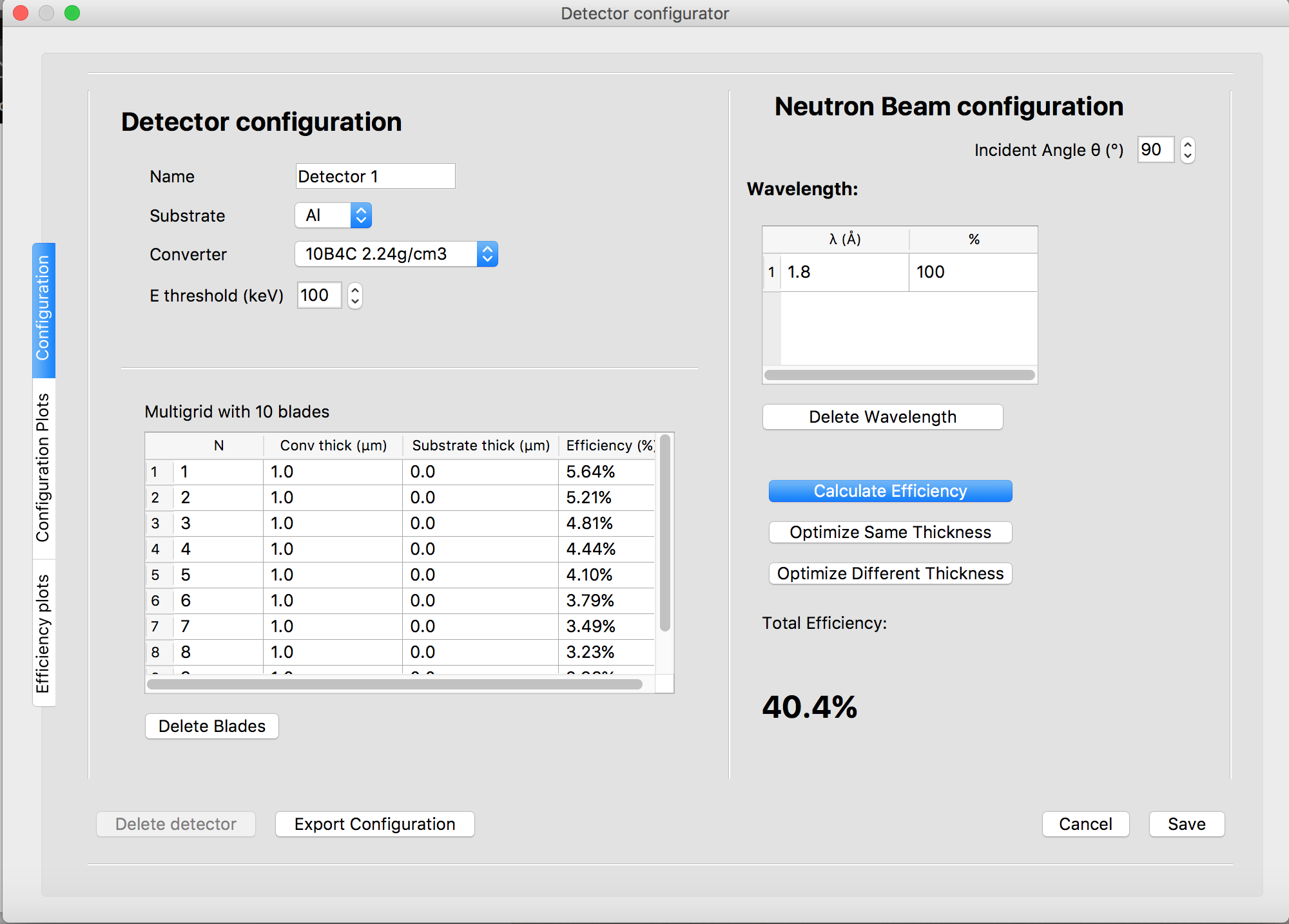}
  \caption{Detector configurator view. This is the window allowing the
    user to define the parameters, display plots and detector efficiency values.}
  \label{guiDetectorConfigurator}
\end{figure}

A graphical user interface (GUI) has been developed for easy access to
the features of the software. The GUI can be executed running the
\textit{launch.py} script, located in the root folder of the
project. This invokes the main window as shown in Fig.~\ref{guimainwindow}.
The detector configuration list is empty to start with. To create a new
configuration click on the \textit{New configuration} button and a new
window pops up (see Fig.~\ref{guiDetectorConfigurator}). Here the
user can configure the neutron detector and neutron beam characteristics. When all the
parameters are set, the user can proceed to calculate the efficiency, get an optimised configuration, export the data from the plots or export the configuration to a \textit{JSON} file. The full list of use cases can be found in the use cases section.
%TODO place this section's figures properly (they are placed in the next)
\FloatBarrier

\section {Use cases}
The following list describes the different functionalities available in the software.

\textit{Create a configuration}: There are different parameters that can be changed with the user input:
%TODO explain the parameters, and change the ones that can't change
%TODO add a multilayer sketch
%TODO add a description of what the user is doing
 name, substrate, converter, and energy threshold. Then the user can \textit{Add Wavelength} and \textit{Add Blades}. The configuration can be saved for later use in this session by clicking \textit{Save}, and can be exported by clicking on \textit{Export Configuration}. 

\begin{figure}[!h]
  \centering
  \includegraphics[width=0.4\columnwidth]{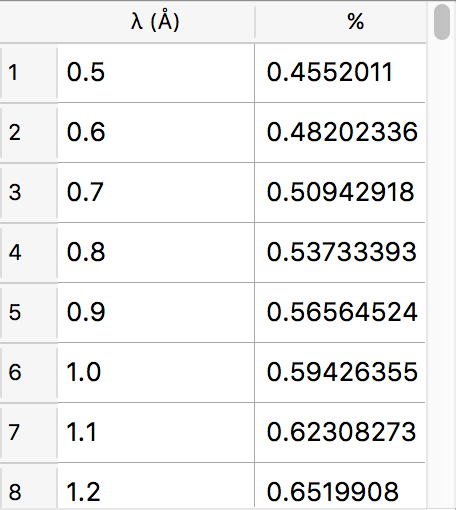}
  \caption{Example of an imported wavelength list with respective weights.}
  \label{wavelist}
\end{figure}

\begin{figure}[!h]
  \centering
  \includegraphics[width=0.9\columnwidth]{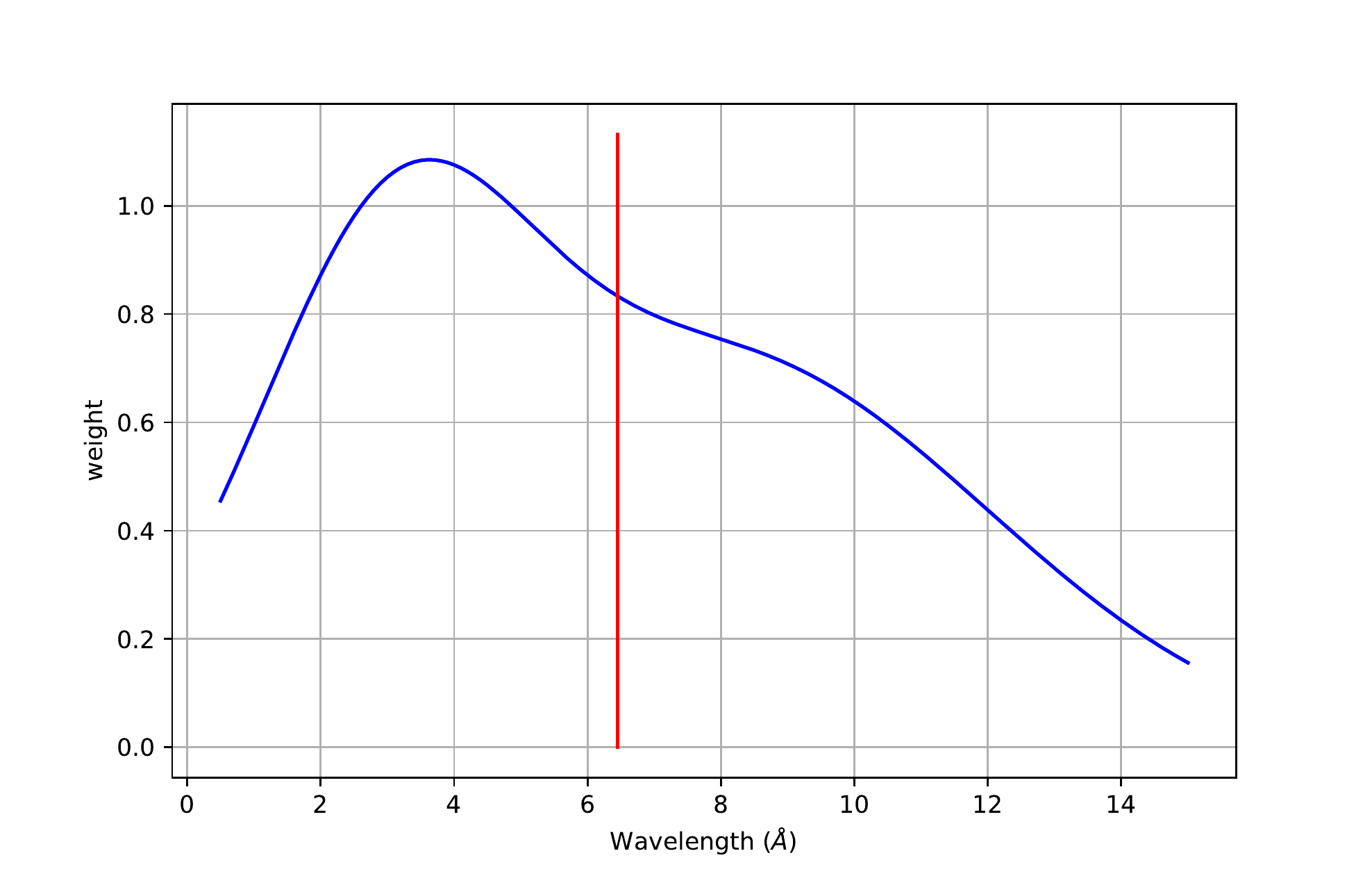}
  \caption{Example of a plot from an imported wavelength list with a double
    Gaussian shape. The red line is at the barycenter of the wavelength distribution.}
  \label{waveplot}
\end{figure}

\textit{Add wavelength}: The user can add wavelength ($\lambda$) in
two ways, either by manually setting a value and its respective ratio
(0--100\%) or via an imported text file. The latter should have a
two-column format as \href{https://github.com/DetectorEfficiencyCalculator/dg_efficiencyCalculator/blob/master/efficiencyCalculator/exports/waves/2gaussdistr.tx}{
  in file 2gaussdistr.txt}. The
tool will display a list of wavelengths and weights (see
Fig.~\ref{wavelist}), and a plot (see Fig.~\ref{waveplot}).

\begin{figure}[!h]
  \centering
  \includegraphics[width=0.7\columnwidth]{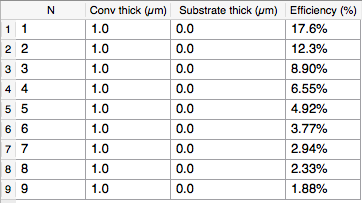}
  \caption{Example of a list of blade thicknesses optimised as a function of depth with the aim of maximizing detector efficiency.}
  \label{opt_table}
\end{figure}

\textit{Add blades}: The user can set the converter thickness and the
number of substrate layers. The GUI provides visualisation aid, i.e.\,a
list with the added blades and plots of the thickness vs.\,the blade
number. The user can set a single layer detector to separately see the
backscattering and transmission efficiency values. To change the blade
configuration the \textit{Delete Blades} button has to be pushed or
the desired value in the blade list can be modified (see Fig.~\ref{opt_table}).   %add figures

\textit{Calculate efficiency}: When the configuration is complete, the
user can calculate the efficiency pressing the  \textit{Calculate
  Efficiency} button or calling the function from the
\textit{scripts.py} file. When the efficiency is calculated, the
window displays the total efficiency for double-coated layers or both
backscattering and transmission values for single layer
configurations. The window displays the respective efficiency of a
blade in a multi-blade configuration. When a multi-blade detector has
all the layers with the same thickness, a plot of the efficiency
vs.\,the blade number and another plot with the total efficiency
vs.\,converter thickness are displayed (see Fig.~\ref{asd}).

\textit{Optimisation}: The list of converter thicknesses of detector can be optimised for maximum
efficiency. The optimisation is done
by calling the \textit{Calculate Efficiency} function changing the
thickness until the maximum is found. In a
multi-blade detector, which has to be optimised for any distribution
of neutron wavelengths or for a single wavelength, each blade has
to hold two converter layers of the same thickness (\cite{piscitelli2013}, page
11). Naturally converter thicknesses on different blades can be
distinct. 

The
most complex operation is the optimisation of a multi-blade detector
with different coating thicknesses and polychromatic wavelength. This
operation could take several minutes for a multi-core CPU but as it
has been demonstrated (\cite{piscitelli2013}, page 16), it is a
sufficient approximation to optimise using the barycenter of the
wavelength distribution. 
\begin{figure}[!t]
  \centering
  \includegraphics[width=0.9\columnwidth]{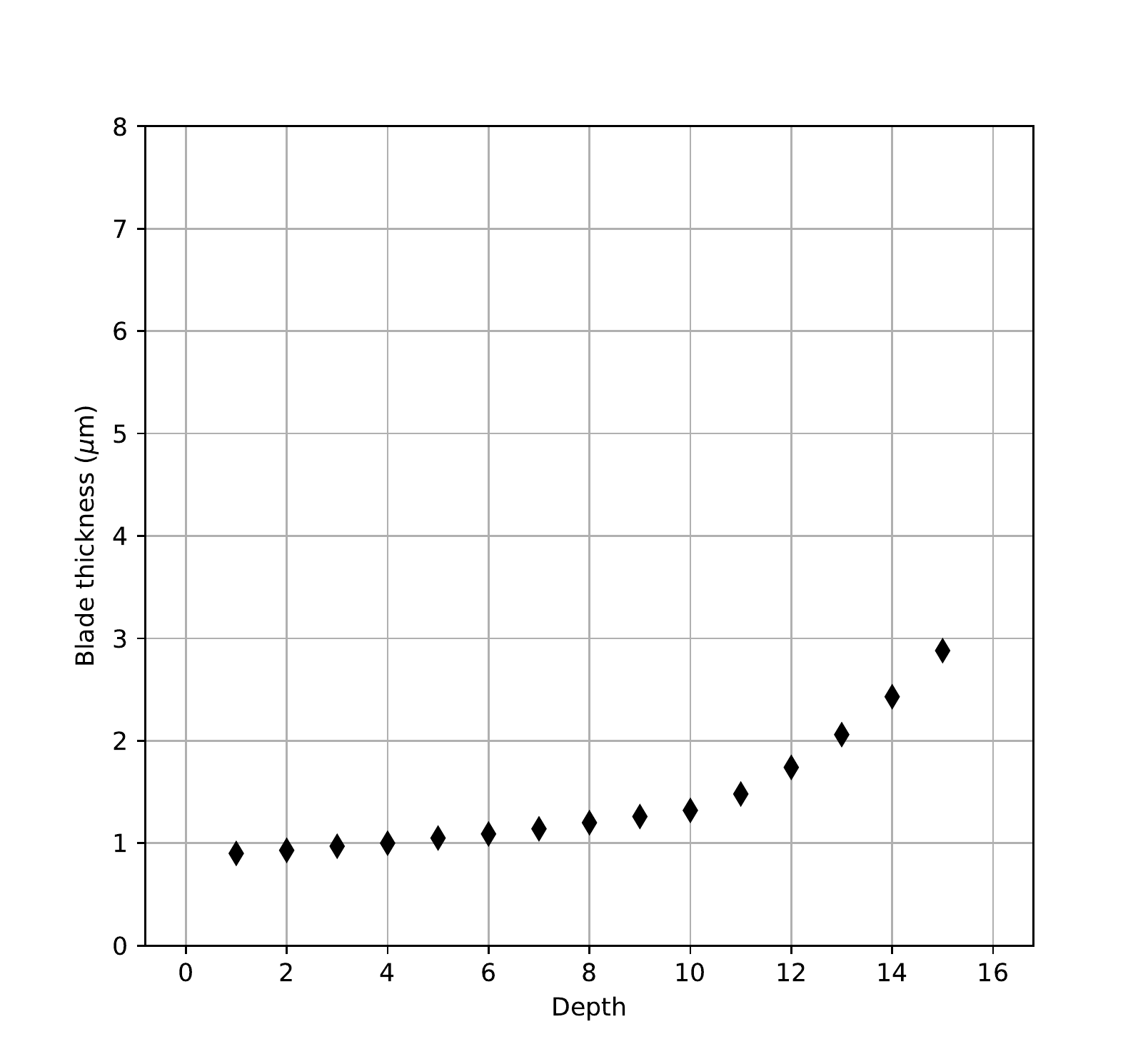}
  \caption{Example of blade thicknesses optimised as a function of depth with the aim of maximizing detector efficiency.}
  \label{opt_plot}
\end{figure}

\textit{Exporting data}: The active neutron detector configuration can
be exported from the detector dialog tab. The configuration is
exported in the \textit{JSON} format. This \textit{JSON} file can be
imported in the main window or used as a parameter for a script function. The plots are exported with the desired format using the \textit{Save} feature in the toolbar under the plots. Data from the plot is exported with the \textit{Export Data} button under some of the plots. This is an example of a configuration exported as a \textit{JSON} file:
%TODO Edit styles to reduce space between lines and put it in a box and cite it from the text, place it elsewhere
\FloatBarrier

\begin{lstlisting}[language=json,firstnumber=1]
{
    "angle": 90, 
    "blades": [
        {
            "backscatter": 4.0, 
            "inclination": 0, 
            "substrate": 0.0, 
            "transmission": 0
        }
    ], 
    "converter": "10B4C 2.24g/cm3", 
    "name": "Detector 1", 
    "single": true, 
    "threshold": 100, 
    "wavelength": [
        {
            "%": 100, 
            "angstrom": 1.8
        }
    ]
}
\end{lstlisting}

\section{Conclusions}
Based on prior analysis of the thin-film neutron detector
characteristics an open source software is developed, in order to calculate
and optimise detector efficiency. The focus of this work is on B10-based neutron
detectors. The Python code produced to this end is accessible via a
GitHub repository for the neutron scattering
community to access, use and modify according to the users' needs. To this end, the calculations are made available as a separate
Python library but can be run via a GUI application as well. Various utilities are offered
that make the software quick to start with and intuitive to use.

The software is available at \\
\href{https://github.com/DetectorEfficiencyCalculator/dg_efficiencyCalculator}{https://github.com/DetectorEfficiencyCalculator/dg\_efficiencyCalculator}
and\\
\href{https://github.com/alvcarmona/neutronDetectorEffFunctions}{https://github.com/alvcarmona/neutronDetectorEffFunctions}. Part
of the DECal functionality is implemented as a web application and can
be found in \href{https://github.com/alvcarmona/efficiencycalculatorweb}{https://github.com/alvcarmona/efficiencycalculatorweb}.

% Further improvements are foreseen and new features will be added in
% future releases, together with a web-based application that will make the software more easily available.

%The tool has proved to be useful for quick calculations within the ESS Detector Group. %TODO
%In the next releases of the software there will be new features like comparison between detectors, inclusion of substrate materials like aluminum, and availability as a web application.

\FloatBarrier

\section*{Acknowledgments}

This work was supported by the EU Horizon 2020 framework, BrightnESS
project 676548. \'Alvaro Carmona Bas\'a\~nez would like to thank the Basque
government and Novia Salcedo for their partial financial support via the Global training
grant.



%% References


%\bibliographystyle{plain}
\bibliographystyle{elsarticle-num}
%\bibliographystyle{model1a-num-names}

%\bibliography{mybibfile}

\begin{thebibliography}{47}
\expandafter\ifx\csname natexlab\endcsname\relax\def\natexlab#1{#1}\fi
\providecommand{\url}[1]{\texttt{#1}}
\providecommand{\href}[2]{#2}
\providecommand{\path}[1]{#1}
\providecommand{\DOIprefix}{doi:}
\providecommand{\ArXivprefix}{arXiv:}
\providecommand{\URLprefix}{URL: }
\providecommand{\Pubmedprefix}{pmid:}
\providecommand{\doi}[1]{\href{http://dx.doi.org/#1}{\path{#1}}}
\providecommand{\Pubmed}[1]{\href{pmid:#1}{\path{#1}}}
\providecommand{\bibinfo}[2]{#2}
\ifx\xfnm\relax \def\xfnm[#1]{\unskip,\space#1}\fi
%Type = Misc
\bibitem[{Peggs et~al.(2013)}]{esstdr}
\bibinfo{author}{S.~Peggs}, et~al., \bibinfo{title}{{ESS Technical Design
  Report, ESS 2013-001}},
  \bibinfo{howpublished}{\url{http://esss.se/scientific-technical-documentation}},
  \bibinfo{year}{2013}.
%Type = Article
\bibitem[{Cooper et~al.(2003)}]{HE3S_cooper}
\bibinfo{author}{R.~Cooper}, et~al., \bibinfo{journal}{A program for neutron
  detector research and development - A White Paper based on a workshop held at
  Oak Ridge National Laboratory, Proceedings of the workshop held at Oak Ridge
  National Laboratory}  (\bibinfo{year}{2003}).
%Type = Article
\bibitem[{Gebauer(2004)}]{HE3S_gebauer}
\bibinfo{author}{B.~Gebauer}, \bibinfo{journal}{Nuclear Instruments and Methods
  in Physics Research Section A} \bibinfo{volume}{535} (\bibinfo{year}{2004})
  \bibinfo{pages}{65 -- 78}. \DOIprefix\doi{10.1016/j.nima.2004.07.266}.
%Type = Article
\bibitem[{Zeitelhack(2012)}]{HE3S_karl}
\bibinfo{author}{K.~Zeitelhack}, \bibinfo{journal}{Neutron News}
  \bibinfo{volume}{23} (\bibinfo{year}{2012}) \bibinfo{pages}{10--13}.
  \DOIprefix\doi{10.1080/10448632.2012.725325}.
%Type = Article
\bibitem[{Kirstein and et~al.(2014)}]{HE3S_kirstein}
\bibinfo{author}{O.~Kirstein}, \bibinfo{author}{et~al.},
  \bibinfo{journal}{arXiv:1411.6194} \bibinfo{volume}{Proceedings of the 23rd
  International Workshop on Vertex Detectors, 15-19 September, Macha Lake, The
  Czech Republic} (\bibinfo{year}{2014}).
%Type = Techreport
\bibitem[{Kouzes(2009)}]{HE3S_kouzes}
\bibinfo{author}{R.~T. Kouzes}, \bibinfo{title}{The $^3$He Supply Problem},
  \bibinfo{type}{Technical Report}, Pacific Northwest National Laboratory,
  Richland, WA, \bibinfo{year}{2009}.
%Type = Article
\bibitem[{Kramer(2011)}]{HE3S_kramer}
\bibinfo{author}{D.~Kramer}, \bibinfo{journal}{Physics Today}
  \bibinfo{volume}{64} (\bibinfo{year}{2011}) \bibinfo{pages}{20--23}.
%Type = Article
\bibitem[{Shea and Morgan(2010)}]{HE3S_shea}
\bibinfo{author}{D.~A. Shea}, \bibinfo{author}{D.~Morgan},
  \bibinfo{journal}{Congressional Research Service}  (\bibinfo{year}{2010}).
%Type = Booklet
\bibitem[{Dianoux and Lander(2013)}]{illblue}
\bibinfo{author}{A.-J. Dianoux}, \bibinfo{author}{G.~Lander},
  \bibinfo{title}{{Neutron Data Booklet}}, \bibinfo{howpublished}{Second
  Edition}, \bibinfo{year}{2013}.
%Type = Article
\bibitem[{H{\"o}glund et~al.(2012)H{\"o}glund, Birch, Andersen, Bigault,
  Buffet, Correa, van Esch, Guerard, Hall-Wilton, Jensen, Khaplanov,
  Piscitelli, Vettier, Vollenberg, and Hultman}]{B4C_carina}
\bibinfo{author}{C.~H{\"o}glund}, \bibinfo{author}{J.~Birch},
  \bibinfo{author}{K.~Andersen}, \bibinfo{author}{T.~Bigault},
  \bibinfo{author}{J.-C. Buffet}, \bibinfo{author}{J.~Correa},
  \bibinfo{author}{P.~van Esch}, \bibinfo{author}{B.~Guerard},
  \bibinfo{author}{R.~Hall-Wilton}, \bibinfo{author}{J.~Jensen},
  \bibinfo{author}{A.~Khaplanov}, \bibinfo{author}{F.~Piscitelli},
  \bibinfo{author}{C.~Vettier}, \bibinfo{author}{W.~Vollenberg},
  \bibinfo{author}{L.~Hultman}, \bibinfo{journal}{Journal of Applied Physics}
  \bibinfo{volume}{111} (\bibinfo{year}{2012}).
  \DOIprefix\doi{10.1063/1.4718573}.
%Type = Article
\bibitem[{Schmidt et~al.(2016)Schmidt, H{\"o}glund, Jensen, Hultman, Birch, and
  Hall-Wilton}]{Schmidt2016}
\bibinfo{author}{S.~Schmidt}, \bibinfo{author}{C.~H{\"o}glund},
  \bibinfo{author}{J.~Jensen}, \bibinfo{author}{L.~Hultman},
  \bibinfo{author}{J.~Birch}, \bibinfo{author}{R.~Hall-Wilton},
  \bibinfo{journal}{Journal of Materials Science} \bibinfo{volume}{51}
  (\bibinfo{year}{2016}) \bibinfo{pages}{10418--10428}.
  \DOIprefix\doi{10.1007/s10853-016-0262-4}.
%Type = Article
\bibitem[{H{\"o}glund et~al.(2015)H{\"o}glund, Zeitelhack, Kudejova, Jensen,
  Greczynski, Lu, Hultman, Birch, and Hall-Wilton}]{HOGLUND201514}
\bibinfo{author}{C.~H{\"o}glund}, \bibinfo{author}{K.~Zeitelhack},
  \bibinfo{author}{P.~Kudejova}, \bibinfo{author}{J.~Jensen},
  \bibinfo{author}{G.~Greczynski}, \bibinfo{author}{J.~Lu},
  \bibinfo{author}{L.~Hultman}, \bibinfo{author}{J.~Birch},
  \bibinfo{author}{R.~Hall-Wilton}, \bibinfo{journal}{Radiation Physics and
  Chemistry} \bibinfo{volume}{113} (\bibinfo{year}{2015})
  \bibinfo{pages}{14--19}. \DOIprefix\doi{10.1016/j.radphyschem.2015.04.006}.
%Type = Article
\bibitem[{Piscitelli(2015)}]{MIO_HERE}
\bibinfo{author}{F.~Piscitelli}, \bibinfo{journal}{The European Physical
  Journal Plus} \bibinfo{volume}{130} (\bibinfo{year}{2015})
  \bibinfo{pages}{1--9}. \DOIprefix\doi{10.1140/epjp/i2015-15027-3}.
%Type = Article
\bibitem[{Piscitelli et~al.(2014)Piscitelli, Buffet, Clergeau, Cuccaro,
  Gu{\'e}rard, Khaplanov, Manna, Rigal, and Esch}]{MIO_MB2014}
\bibinfo{author}{F.~Piscitelli}, \bibinfo{author}{J.~C. Buffet},
  \bibinfo{author}{J.~F. Clergeau}, \bibinfo{author}{S.~Cuccaro},
  \bibinfo{author}{B.~Gu{\'e}rard}, \bibinfo{author}{A.~Khaplanov},
  \bibinfo{author}{Q.~L. Manna}, \bibinfo{author}{J.~M. Rigal},
  \bibinfo{author}{P.~V. Esch}, \bibinfo{journal}{Journal of Instrumentation}
  \bibinfo{volume}{9} (\bibinfo{year}{2014}) \bibinfo{pages}{P03007}.
%Type = Article
\bibitem[{Piscitelli et~al.(2017)Piscitelli, Messi, Anastasopoulos, Bry{\'s},
  Chicken, Dian, Fuzi, H{\"o}glund, Kiss, Orban, Pazmandi, Robinson, Rosta,
  Schmidt, Varga, Zsiros, and Hall-Wilton}]{MIO_MB2017}
\bibinfo{author}{F.~Piscitelli}, \bibinfo{author}{F.~Messi},
  \bibinfo{author}{M.~Anastasopoulos}, \bibinfo{author}{T.~Bry{\'s}},
  \bibinfo{author}{F.~Chicken}, \bibinfo{author}{E.~Dian},
  \bibinfo{author}{J.~Fuzi}, \bibinfo{author}{C.~H{\"o}glund},
  \bibinfo{author}{G.~Kiss}, \bibinfo{author}{J.~Orban},
  \bibinfo{author}{P.~Pazmandi}, \bibinfo{author}{L.~Robinson},
  \bibinfo{author}{L.~Rosta}, \bibinfo{author}{S.~Schmidt},
  \bibinfo{author}{D.~Varga}, \bibinfo{author}{T.~Zsiros},
  \bibinfo{author}{R.~Hall-Wilton}, \bibinfo{journal}{Journal of
  Instrumentation} \bibinfo{volume}{12} (\bibinfo{year}{2017})
  \bibinfo{pages}{P03013}.
%Type = Article
\bibitem[{Henske et~al.(2012)Henske, Klein, K{\"o}hli, Lennert, Modzel,
  Schmidt, and Schmidt}]{DET_jalousie}
\bibinfo{author}{M.~Henske}, \bibinfo{author}{M.~Klein},
  \bibinfo{author}{M.~K{\"o}hli}, \bibinfo{author}{P.~Lennert},
  \bibinfo{author}{G.~Modzel}, \bibinfo{author}{C.~Schmidt},
  \bibinfo{author}{U.~Schmidt}, \bibinfo{journal}{Nuclear Instruments and
  Methods in Physics Research Section A} \bibinfo{volume}{686}
  (\bibinfo{year}{2012}) \bibinfo{pages}{151 -- 155}.
  \DOIprefix\doi{10.1016/j.nima.2012.05.075}.
%Type = Article
\bibitem[{Modzel et~al.(2014)Modzel, Henske, Houben, Klein, K{\"o}hli, Lennert,
  Meven, Schmidt, Schmidt, and Schweika}]{DET_Jalousie3}
\bibinfo{author}{G.~Modzel}, \bibinfo{author}{M.~Henske},
  \bibinfo{author}{A.~Houben}, \bibinfo{author}{M.~Klein},
  \bibinfo{author}{M.~K{\"o}hli}, \bibinfo{author}{P.~Lennert},
  \bibinfo{author}{M.~Meven}, \bibinfo{author}{C.~Schmidt},
  \bibinfo{author}{U.~Schmidt}, \bibinfo{author}{W.~Schweika},
  \bibinfo{journal}{Nuclear Instruments and Methods in Physics Research Section
  A} \bibinfo{volume}{743} (\bibinfo{year}{2014}) \bibinfo{pages}{90 -- 95}.
  \DOIprefix\doi{10.1016/j.nima.2014.01.007}.
%Type = Article
\bibitem[{Kampmann(2012{\natexlab{a}})}]{DET_kampmannA1CLDp}
\bibinfo{author}{R.~Kampmann}, \bibinfo{journal}{IEEE NSS, Anaheim}
  \bibinfo{volume}{N1-109} (\bibinfo{year}{2012}{\natexlab{a}}).
%Type = Article
\bibitem[{Kampmann(2012{\natexlab{b}})}]{DET_kampmannA1CLDp2}
\bibinfo{author}{R.~Kampmann}, \bibinfo{journal}{First Tests of Novel Neutron
  Detectors with Thin Conversion Layers in Inclined Geometry at {REFSANS}, Oral
  presentation at the Second International $^{10}$B-BF$_3$ Detectors Workshop,
  13-14 March, ILL, Grenoble}  (\bibinfo{year}{2012}{\natexlab{b}}). \URLprefix
  \url{https://www.ill.eu/10bbf3}.
%Type = Article
\bibitem[{Cippo et~al.(2015)Cippo, Croci, Muraro, Menelle, Albani, Cavenago,
  Cazzaniga, Claps, Grosso, Murtas, Rebai, Tardocchi, and
  Gorini}]{MPGD_CrociRate}
\bibinfo{author}{E.~P. Cippo}, \bibinfo{author}{G.~Croci},
  \bibinfo{author}{A.~Muraro}, \bibinfo{author}{A.~Menelle},
  \bibinfo{author}{G.~Albani}, \bibinfo{author}{M.~Cavenago},
  \bibinfo{author}{C.~Cazzaniga}, \bibinfo{author}{G.~Claps},
  \bibinfo{author}{G.~Grosso}, \bibinfo{author}{F.~Murtas},
  \bibinfo{author}{M.~Rebai}, \bibinfo{author}{M.~Tardocchi},
  \bibinfo{author}{G.~Gorini}, \bibinfo{journal}{Journal of Instrumentation}
  \bibinfo{volume}{10} (\bibinfo{year}{2015}) \bibinfo{pages}{P10003}.
%Type = Article
\bibitem[{Croci and et~al.(2014)}]{MPGD_GEMcroci}
\bibinfo{author}{G.~Croci}, \bibinfo{author}{et~al.}, \bibinfo{journal}{EPL}
  \bibinfo{volume}{107} (\bibinfo{year}{2014}) \bibinfo{pages}{12001}.
  \DOIprefix\doi{10.1209/0295-5075/107/12001}.
%Type = Article
\bibitem[{Anastasopoulos et~al.(2017)Anastasopoulos, Bebb, Berry, Birch,
  Bry{\'s}, Buffet, Clergeau, Deen, Ehlers, van Esch, Everett, Guerard,
  Hall-Wilton, Herwig, Hultman, H{\"o}glund, Iruretagoiena, Issa, Jensen,
  Khaplanov, Kirstein, Higuera, Piscitelli, Robinson, Schmidt, and
  Stefanescu}]{MG_2017}
\bibinfo{author}{M.~Anastasopoulos}, \bibinfo{author}{R.~Bebb},
  \bibinfo{author}{K.~Berry}, \bibinfo{author}{J.~Birch},
  \bibinfo{author}{T.~Bry{\'s}}, \bibinfo{author}{J.-C. Buffet},
  \bibinfo{author}{J.-F. Clergeau}, \bibinfo{author}{P.~Deen},
  \bibinfo{author}{G.~Ehlers}, \bibinfo{author}{P.~van Esch},
  \bibinfo{author}{S.~Everett}, \bibinfo{author}{B.~Guerard},
  \bibinfo{author}{R.~Hall-Wilton}, \bibinfo{author}{K.~Herwig},
  \bibinfo{author}{L.~Hultman}, \bibinfo{author}{C.~H{\"o}glund},
  \bibinfo{author}{I.~Iruretagoiena}, \bibinfo{author}{F.~Issa},
  \bibinfo{author}{J.~Jensen}, \bibinfo{author}{A.~Khaplanov},
  \bibinfo{author}{O.~Kirstein}, \bibinfo{author}{I.~L. Higuera},
  \bibinfo{author}{F.~Piscitelli}, \bibinfo{author}{L.~Robinson},
  \bibinfo{author}{S.~Schmidt}, \bibinfo{author}{I.~Stefanescu},
  \bibinfo{journal}{Journal of Instrumentation} \bibinfo{volume}{12}
  (\bibinfo{year}{2017}) \bibinfo{pages}{P04030}.
%Type = Article
\bibitem[{Andersen et~al.(2013)Andersen, Bigault, Birch, Buffet, Correa,
  Hall-Wilton, Hultman, H{\"o}glund, Gu{\'e}rard, Jensen, Khaplanov, Kirstein,
  Piscitelli, Esch, and Vettier}]{MG_andersen}
\bibinfo{author}{K.~Andersen}, \bibinfo{author}{T.~Bigault},
  \bibinfo{author}{J.~Birch}, \bibinfo{author}{J.~Buffet},
  \bibinfo{author}{J.~Correa}, \bibinfo{author}{R.~Hall-Wilton},
  \bibinfo{author}{L.~Hultman}, \bibinfo{author}{C.~H{\"o}glund},
  \bibinfo{author}{B.~Gu{\'e}rard}, \bibinfo{author}{J.~Jensen},
  \bibinfo{author}{A.~Khaplanov}, \bibinfo{author}{O.~Kirstein},
  \bibinfo{author}{F.~Piscitelli}, \bibinfo{author}{P.~V. Esch},
  \bibinfo{author}{C.~Vettier}, \bibinfo{journal}{Nuclear Instruments and
  Methods in Physics Research Section A} \bibinfo{volume}{720}
  (\bibinfo{year}{2013}) \bibinfo{pages}{116 -- 121}.
  \DOIprefix\doi{10.1016/j.nima.2012.12.021}.
%Type = Article
\bibitem[{Birch et~al.(2014)Birch, Buffet, Clergeau, Correa, van Esch,
  Ferraton, Guerard, Halbwachs, Hall-Wilton, Hultman, H{\"o}glund, Khaplanov,
  Koza, Piscitelli, and Zbiri}]{MG_IN6tests}
\bibinfo{author}{J.~Birch}, \bibinfo{author}{J.-C. Buffet},
  \bibinfo{author}{J.-F. Clergeau}, \bibinfo{author}{J.~Correa},
  \bibinfo{author}{P.~van Esch}, \bibinfo{author}{M.~Ferraton},
  \bibinfo{author}{B.~Guerard}, \bibinfo{author}{J.~Halbwachs},
  \bibinfo{author}{R.~Hall-Wilton}, \bibinfo{author}{L.~Hultman},
  \bibinfo{author}{C.~H{\"o}glund}, \bibinfo{author}{A.~Khaplanov},
  \bibinfo{author}{M.~Koza}, \bibinfo{author}{F.~Piscitelli},
  \bibinfo{author}{M.~Zbiri}, \bibinfo{journal}{Journal of Physics: Conference
  Series} \bibinfo{volume}{528} (\bibinfo{year}{2014}) \bibinfo{pages}{012040}.
%Type = Article
\bibitem[{Birch et~al.(2013)Birch, Buffet, Correa, van Esch, Gu{\'e}rard,
  Hall-Wilton, H{\"o}glund, Hultman, Khaplanov, and Piscitelli}]{MG_joni}
\bibinfo{author}{J.~Birch}, \bibinfo{author}{J.~C. Buffet},
  \bibinfo{author}{J.~Correa}, \bibinfo{author}{P.~van Esch},
  \bibinfo{author}{B.~Gu{\'e}rard}, \bibinfo{author}{R.~Hall-Wilton},
  \bibinfo{author}{C.~H{\"o}glund}, \bibinfo{author}{L.~Hultman},
  \bibinfo{author}{A.~Khaplanov}, \bibinfo{author}{F.~Piscitelli},
  \bibinfo{journal}{IEEE Transactions on Nuclear Science} \bibinfo{volume}{60}
  (\bibinfo{year}{2013}) \bibinfo{pages}{871--878}.
  \DOIprefix\doi{10.1109/TNS.2012.2227798}.
%Type = Misc
\bibitem[{Guerard and Buffet(2011)}]{MG_patent}
\bibinfo{author}{B.~Guerard}, \bibinfo{author}{J.~Buffet},
  \bibinfo{title}{Ionizing radiation detector - patent no. 20110215251.},
  \bibinfo{year}{2011}. \URLprefix
  \url{https://www.google.com/patents/US20110215251}, \bibinfo{note}{{US}
  Patent App. 13/038,915}.
%Type = Article
\bibitem[{Klein and Schmidt(2011)}]{MPGD_KleinCASCADE}
\bibinfo{author}{M.~Klein}, \bibinfo{author}{C.~J. Schmidt},
  \bibinfo{journal}{Nuclear Instruments and Methods in Physics Research Section
  A} \bibinfo{volume}{628} (\bibinfo{year}{2011}) \bibinfo{pages}{9--18}.
  \DOIprefix\doi{10.1016/j.nima.2010.06.278}, \bibinfo{note}{\{VCI\}
  2010Proceedings of the 12th International Vienna Conference on
  Instrumentation}.
%Type = Phdthesis
\bibitem[{Klein(2000)}]{kleinphd}
\bibinfo{author}{M.~Klein}, \bibinfo{title}{{Experimente zur Quantenmechanik
  mit ultrakalten Neutronen und Entwicklung eines neuen Detektors zum
  ortsaufgeloesten Nachweis von thermischen Neutronen auf grossen
  Fl{\"a}chen}}, Ph.D. thesis, University of Heidelberg, \bibinfo{year}{2000}.
%Type = Article
\bibitem[{Stefanescu et~al.(2013{\natexlab{a}})Stefanescu, Abdullahi, Birch,
  Defendi, Hall-Wilton, H{\"o}glund, Hultman, Seiler, and
  Zeitelhack}]{DET_stefanescu1}
\bibinfo{author}{I.~Stefanescu}, \bibinfo{author}{Y.~Abdullahi},
  \bibinfo{author}{J.~Birch}, \bibinfo{author}{I.~Defendi},
  \bibinfo{author}{R.~Hall-Wilton}, \bibinfo{author}{C.~H{\"o}glund},
  \bibinfo{author}{L.~Hultman}, \bibinfo{author}{D.~Seiler},
  \bibinfo{author}{K.~Zeitelhack}, \bibinfo{journal}{Nuclear Instruments and
  Methods in Physics Research Section A} \bibinfo{volume}{727}
  (\bibinfo{year}{2013}{\natexlab{a}}) \bibinfo{pages}{109 -- 125}.
  \DOIprefix\doi{10.1016/j.nima.2013.06.003}.
%Type = Article
\bibitem[{Stefanescu et~al.(2013{\natexlab{b}})Stefanescu, Abdullahi, Birch,
  Defendi, Hall-Wilton, H{\"o}glund, Hultman, Zee, and
  Zeitelhack}]{DET_stefanescu2}
\bibinfo{author}{I.~Stefanescu}, \bibinfo{author}{Y.~Abdullahi},
  \bibinfo{author}{J.~Birch}, \bibinfo{author}{I.~Defendi},
  \bibinfo{author}{R.~Hall-Wilton}, \bibinfo{author}{C.~H{\"o}glund},
  \bibinfo{author}{L.~Hultman}, \bibinfo{author}{M.~Zee},
  \bibinfo{author}{K.~Zeitelhack}, \bibinfo{journal}{Journal of
  Instrumentation} \bibinfo{volume}{8} (\bibinfo{year}{2013}{\natexlab{b}})
  \bibinfo{pages}{P12003}.
%Type = Article
\bibitem[{Pfeiffer et~al.(2015)Pfeiffer, Resnati, Birch, Hall-Wilton, Höglund,
  Hultman, Iakovidis, Oliveri, Oksanen, Ropelewski, and
  Thuiner}]{MPGD_pfeiffer2015}
\bibinfo{author}{D.~Pfeiffer}, \bibinfo{author}{F.~Resnati},
  \bibinfo{author}{J.~Birch}, \bibinfo{author}{R.~Hall-Wilton},
  \bibinfo{author}{C.~Höglund}, \bibinfo{author}{L.~Hultman},
  \bibinfo{author}{G.~Iakovidis}, \bibinfo{author}{E.~Oliveri},
  \bibinfo{author}{E.~Oksanen}, \bibinfo{author}{L.~Ropelewski},
  \bibinfo{author}{P.~Thuiner}, \bibinfo{journal}{Journal of Instrumentation}
  \bibinfo{volume}{10} (\bibinfo{year}{2015}) \bibinfo{pages}{P04004}.
%Type = Inproceedings
\bibitem[{Lacy et~al.(2009)Lacy, Sun, Martin, Athanasiades, and
  Lyons}]{STRAW_lacy2}
\bibinfo{author}{J.~L. Lacy}, \bibinfo{author}{L.~Sun}, \bibinfo{author}{C.~S.
  Martin}, \bibinfo{author}{A.~Athanasiades}, \bibinfo{author}{T.~D. Lyons},
  in: \bibinfo{booktitle}{2009 IEEE Nuclear Science Symposium Conference Record
  (NSS/MIC)}, pp. \bibinfo{pages}{1117--1121}.
  \DOIprefix\doi{10.1109/NSSMIC.2009.5402421}.
%Type = Article
\bibitem[{Lacy et~al.(2002)Lacy, Athanasiades, Shehad, Austin, and
  Martin}]{STRAW_lacy2002}
\bibinfo{author}{J.~L. Lacy}, \bibinfo{author}{A.~Athanasiades},
  \bibinfo{author}{N.~N. Shehad}, \bibinfo{author}{R.~A. Austin},
  \bibinfo{author}{C.~S. Martin}, \bibinfo{journal}{IEEE NSS, Norfolk}
  \bibinfo{volume}{1} (\bibinfo{year}{2002}) \bibinfo{pages}{392--396}.
%Type = Article
\bibitem[{Lacy(2006)}]{STRAW_lacy2006}
\bibinfo{author}{J.~Lacy}, \bibinfo{journal}{IEEE Nuclear Science Symposium
  Conference Record} \bibinfo{volume}{1} (\bibinfo{year}{2006})
  \bibinfo{pages}{20--26}.
%Type = Article
\bibitem[{Lacy et~al.(2011)Lacy, Athanasiades, Sun, Martin, Lyons, Foss, and
  Haygood}]{STRAW_lacy2011}
\bibinfo{author}{J.~L. Lacy}, \bibinfo{author}{A.~Athanasiades},
  \bibinfo{author}{L.~Sun}, \bibinfo{author}{C.~S. Martin},
  \bibinfo{author}{T.~D. Lyons}, \bibinfo{author}{M.~A. Foss},
  \bibinfo{author}{H.~B. Haygood}, \bibinfo{journal}{Nuclear Instruments and
  Methods in Physics Research Section A} \bibinfo{volume}{652}
  (\bibinfo{year}{2011}) \bibinfo{pages}{359 -- 363}.
  \DOIprefix\doi{10.1016/j.nima.2010.09.011}, \bibinfo{note}{symposium on
  Radiation Measurements and Applications (SORMA) \{XII\} 2010}.
%Type = Article
\bibitem[{Lacy et~al.(2012)}]{STRAW_lacy2012}
\bibinfo{author}{J.~L. Lacy}, et~al., \bibinfo{journal}{IEEE NSS, Anaheim}
  \bibinfo{volume}{He-1-6} (\bibinfo{year}{2012}).
%Type = Article
\bibitem[{Lacy et~al.(2013)Lacy, Athanasiades, Martin, Sun, and
  Vazquez-Flores}]{STRAW_lacy2013}
\bibinfo{author}{J.~L. Lacy}, \bibinfo{author}{A.~Athanasiades},
  \bibinfo{author}{C.~S. Martin}, \bibinfo{author}{L.~Sun},
  \bibinfo{author}{G.~L. Vazquez-Flores}, \bibinfo{journal}{IEEE Trans. Nucl.
  Sci.} \bibinfo{volume}{60} (\bibinfo{year}{2013})
  \bibinfo{pages}{1140--1146}.
%Type = Article
\bibitem[{McGregor et~al.(2003)McGregor, Hammig, Yang, Gersch, and
  Klann}]{mcgregor}
\bibinfo{author}{D.~McGregor}, \bibinfo{author}{M.~Hammig},
  \bibinfo{author}{Y.-H. Yang}, \bibinfo{author}{H.~Gersch},
  \bibinfo{author}{R.~Klann}, \bibinfo{journal}{Nuclear Instruments and Methods
  in Physics Research Section A} \bibinfo{volume}{500} (\bibinfo{year}{2003})
  \bibinfo{pages}{272 -- 308}. \DOIprefix\doi{10.1016/S0168-9002(02)02078-8}.
%Type = Article
\bibitem[{Piscitelli and Esch(2013)}]{piscitelli2013}
\bibinfo{author}{F.~Piscitelli}, \bibinfo{author}{P.~V. Esch},
  \bibinfo{journal}{Journal of Instrumentation} \bibinfo{volume}{8}
  (\bibinfo{year}{2013}) \bibinfo{pages}{P04020}.
%Type = Article
\bibitem[{Kanaki et~al.(2017)Kanaki, Kittelmann, Cai, Klinkby, Knudsen,
  Willendrup, and Hall-Wilton}]{icnskelly}
\bibinfo{author}{K.~Kanaki}, \bibinfo{author}{T.~Kittelmann},
  \bibinfo{author}{X.~X. Cai}, \bibinfo{author}{E.~Klinkby},
  \bibinfo{author}{E.~B. Knudsen}, \bibinfo{author}{P.~Willendrup},
  \bibinfo{author}{R.~Hall-Wilton}, \bibinfo{journal}{Arxiv preprint hep-th}
  (\bibinfo{year}{2017}). \href{http://arxiv.org/abs/1708.02135}{\tt
  arXiv:1708.02135}.
%Type = Article
\bibitem[{Piscitelli and Esch(2013)}]{piscitelli2013bis}
\bibinfo{author}{F.~Piscitelli}, \bibinfo{author}{P.~V. Esch},
  \bibinfo{journal}{Journal of Instrumentation} \bibinfo{volume}{8}
  (\bibinfo{year}{2013}) \bibinfo{pages}{P04020}.
%Type = Phdthesis
\bibitem[{Piscitelli(3133)}]{FPThesis}
\bibinfo{author}{F.~Piscitelli}, \bibinfo{title}{Boron-10 layers, Neutron
  Reflectometry and Thermal Neutron Gaseous Detectors}, Ph.D. thesis, Institut
  Laue-Langevin and University of Perugia, \bibinfo{year}{2014 -
  arXiv:1406.3133}.
%Type = Webpage
\bibitem[{scipy(????)}]{scipy}
\bibinfo{author}{scipy}, ????
\newblock \URLprefix \url{https://www.scipy.org/}.
%Type = Webpage
\bibitem[{numpy(????)}]{numpy}
\bibinfo{author}{numpy}, ????
\newblock \URLprefix \url{http://www.numpy.org/}.
%Type = Webpage
\bibitem[{matplotlib(????)}]{matplotlib}
\bibinfo{author}{matplotlib}, ????
\newblock \URLprefix \url{https://matplotlib.org/}.
%Type = Webpage
\bibitem[{tutorialspoint(????)}]{builderp}
\bibinfo{author}{tutorialspoint}, ????
\newblock \URLprefix
  \url{https://www.tutorialspoint.com/design_pattern/builder_pattern.htm}.
%Type = Webpage
\bibitem[{Ziegler(????)}]{SRIM}
\bibinfo{author}{J.~F. Ziegler}, ????
\newblock \URLprefix \url{http://www.srim.org/}.

\end{thebibliography}



\end{document}